\documentclass[shortnote,twocolumn]{jpsj3}
\usepackage{txfonts}
\usepackage{amsmath,amssymb,bm}
\usepackage[colorlinks=true,linkcolor=blue,citecolor=blue,urlcolor=blue]{hyperref}
\newcommand{\ii}{\mathrm{i}}
\newcommand{\Pf}{\operatorname{Pf}}
\newcommand{\PG}{P_{\mathrm G}}
\newcommand{\ketvac}{\lvert 0\rangle}
\usepackage{balance}
\makeatletter
\if@twocolumn\@lastpagebalancingfalse\fi
\providecommand{\ext@figure}{lof}
\providecommand{\ext@table}{lot}
\patchcmd{\thebibliography}{\footnotesize}{\fontsize{7.5pt}{8pt}\selectfont}
\makeatother

\title{Embedding Paired Free-Fermion Gaussian States into Gutzwiller-Projected Bardeen--Cooper--Schrieffer Wave Functions}
\author{Ryui Kaneko\thanks{E-mail address: ryuikaneko@sophia.ac.jp}}
\inst{Physics Division, Sophia University, Chiyoda, Tokyo 102-8554, Japan}
\abst{%
Gutzwiller-projected Bardeen--Cooper--Schrieffer (BCS) wave functions of Abrikosov fermions are widely used to describe quantum many-body states.
Taking the one-dimensional transverse-field Ising model as an example,
we exactly embed any even-parity spinless fermionic Gaussian state representable as a paired exponential in the chosen particle basis
into a projected BCS state of spinful Abrikosov fermions.
The construction provides controlled initial states for variational Monte Carlo studies of nonintegrable models.
}

\begin{document}

\maketitle

Variational wave functions provide concrete representations of quantum phases.
Gutzwiller-projected Bardeen--Cooper--Schrieffer (BCS) states describe a wide range of strongly correlated phases~\cite{anderson1987,gros1988,lee2006}
and, in special cases, become exact ground states of quantum spin models~\cite{haldane1988,shastry1988,affleck1987,liu2012,tu2013,kitaev2006,misawa2026}.
Nevertheless,
an explicit Abrikosov-fermion projected BCS representation has not been established even for the well-known one-dimensional (1D) transverse-field Ising (TFI) model,
although its ground state can be expressed exactly as a Gaussian state of Jordan--Wigner (JW) fermions~\cite{pfeuty1970}.
Here we provide such a construction by introducing an auxiliary Pfaffian that evaluates to unity for every spin configuration.

We consider the 1D TFI model on a periodic chain of even $L$ sites:
\begin{align}
 \label{eq:tfim}
 H_{\mathrm{TFI}}
 =
 -J\sum_{j=1}^{L}\sigma_j^x\sigma_{j+1}^x
 -h\sum_{j=1}^{L}\sigma_j^z,
 \quad
 g = h/J,
 \quad
 J>0,
\end{align}
where $J$ is the coupling constant and $h$ is the strength of the transverse field.
The operators $\sigma_j^x$ and $\sigma_j^z$ are Pauli matrices.
After the JW transformation defined by
$\sigma_j^z=1-2a_j^\dagger a_j=1-2n_j$ and
$\sigma_j^x=\left[\prod_{\ell<j}(1-2n_\ell)\right](a_j^\dagger+a_j)$,
spins are replaced by noninteracting fermions under antiperiodic boundary conditions for the finite-system even-parity ground state.
Identifying a down spin with an occupied site, the ground state, up to an overall normalization, is given by
\begin{align}
 \label{eq:jwbcs}
 |\Psi_{\mathrm{JW}}\rangle
 &= \exp\!\left(\sum_{i,j=1}^{L} \frac{\psi_{i,j}}{2}
 a_i^\dagger a_j^\dagger\right)\ketvac,
 \quad
 \psi_{i,j}=\frac1L\sum_k e^{\ii k(j-i)}\psi_k,
\\
 \label{eq:phik}
 \psi_k&=\ii\frac{\sin k}{g-\cos k+\epsilon_k},
 \quad
 \epsilon_k=\sqrt{1+g^2-2g\cos k},
\end{align}
with the momenta defined by $k=2\pi(m+1/2)/L$ for $m=0,1,\ldots,L-1$.
For a given
$|D\rangle = a_{d_1}^\dagger a_{d_2}^\dagger \cdots a_{d_{2m}}^\dagger\ketvac$,
with the set of down-spin sites $D=\{d_1,d_2,\dots,d_{2m}\}$
($d_1<d_2<\cdots<d_{2m}$),
the overlap between $|D\rangle$ and the ground state is
\begin{align}
 \label{eq:jwpf}
 \Psi_{\mathrm{JW}}(D)
 =
 \frac{1}{2^m m!}\sum_{\tau\in \mathfrak{S}_{2m}}\operatorname{sgn}(\tau)\prod_{j=1}^{m}\psi_{d_{\tau(2j-1)},d_{\tau(2j)}}
 =
 \Pf \bm{\psi}_D,
\end{align}
where $\bm{\psi}_D$ is a $2m\times2m$ antisymmetric matrix with elements $\psi_{d_a,d_b}$ for $a,b=1,2,\dots,2m$,
$\mathfrak{S}_{2m}$ is the set of all permutations of $1,2,\dots,2m$,
and $\Pf$ represents a Pfaffian.

This wave function $\Psi_{\mathrm{JW}}(D)$ can be embedded into the projected BCS representation of Abrikosov fermions, satisfying
$|\uparrow\rangle_j = c_{j,\uparrow}^\dagger\ketvac$ and
$|\downarrow\rangle_j = c_{j,\downarrow}^\dagger\ketvac$, with
$n_{j,\uparrow}+n_{j,\downarrow}=1$.
Let us define the wave function
\begin{align}
 \label{eq:pgbcs}
 |\Phi[F]\rangle = \PG\exp\!\left(
 \sum_{i,j=1}^{L} \sum_{\alpha,\beta \in \{\uparrow,\downarrow\}} \frac{F_{i,j}^{\alpha,\beta}}{2}
 c_{i,\alpha}^\dagger c_{j,\beta}^\dagger\right)\ketvac,
\end{align}
where $F_{i,j}^{\alpha,\beta}=-F_{j,i}^{\beta,\alpha}$ is a pairing amplitude
and $\PG = \prod_i \delta_{n_{i,\uparrow}+n_{i,\downarrow},1}$ imposes one fermion per site.
We choose
\begin{align}
 \label{eq:choice}
 F_{i,j}^{\downarrow,\downarrow} = \psi_{i,j},\quad
 F_{i,j}^{\uparrow,\uparrow} = s_{i,j},\quad
 F_{i,j}^{\uparrow,\downarrow} = -F_{j,i}^{\downarrow,\uparrow} = 0,
\end{align}
with $s_{i,j}=-\operatorname{sgn}(i-j)=1,-1,0$ for $i<j$, $i>j$, and $i=j$, respectively.
Let $U=\{u_1,u_2,\dots,u_{2p}\}$
($u_1<u_2<\cdots<u_{2p}$, $2p=L-2m$)
be the up-spin set, complementary to the down-spin set $D$.
We define the fermionic representative of each physical spin configuration by
\begin{align}
 \label{eq:basis_order}
 |x\rangle =
 c_{u_1,\uparrow}^{\dagger}\cdots
 c_{u_{2p},\uparrow}^{\dagger}
 c_{d_1,\downarrow}^{\dagger}\cdots
 c_{d_{2m},\downarrow}^{\dagger}\ketvac.
\end{align}
Here, all up-spin operators precede the down-spin operators, with each set ordered by increasing site index.
Equation~(\ref{eq:basis_order}) defines the spin-to-fermion basis mapping,
rather than a reordering of a separately defined site-ordered Fock basis.
A different ordering transforms both the basis states and the pairing matrix consistently, leaving the physical spin state unchanged.
Because $|x\rangle$ already contains exactly one fermion per site,
$\langle x|\PG=\langle x|$ holds. Hence,
\begin{align}
 \label{eq:amplitude_explicit}
 \Phi(x)
 &=
 \langle x|\Phi[F]\rangle
 =
 \langle 0|
 c_{d_{2m},\downarrow}\cdots c_{d_1,\downarrow}
 c_{u_{2p},\uparrow}\cdots c_{u_1,\uparrow}
\nonumber
\\
 &
 \times
 \exp\left[
 \sum_{i,j=1}^{L} \frac{s_{i,j}}{2}
 c_{i,\uparrow}^{\dagger}c_{j,\uparrow}^{\dagger}
 +
 \sum_{i,j=1}^{L} \frac{\psi_{i,j}}{2}
 c_{i,\downarrow}^{\dagger}c_{j,\downarrow}^{\dagger}
 \right]\ketvac .
\end{align}
Only the term containing $p$ up-spin pairs and $m$ down-spin pairs contributes to the amplitude,
and the corresponding principal submatrix of the pairing matrix is block diagonal:
\begin{align}
 \bm{F}_x =
 \begin{pmatrix}
  \bm{s}_U & 0\\
  0 & \bm{\psi}_D
 \end{pmatrix},
 \quad
 (\bm{s}_U)_{a,b}=s_{u_a,u_b},\quad
 (\bm{\psi}_D)_{a,b}=\psi_{d_a,d_b}.
\end{align}
Since the Pfaffian of a block diagonal matrix is the product of the Pfaffians of its blocks,
$
 \Phi(x)
 =\Pf \bm{F}_x
 =\Pf \bm{s}_U \cdot \Pf \bm{\psi}_D
$
holds.
No additional permutation sign appears because the operator ordering in Eq.~(\ref{eq:basis_order}) matches the block ordering in $\bm{F}_x$.

Since $u_1<\cdots<u_{2p}$, the restricted antisymmetric matrix $\bm{s}_U$
has matrix elements $(\bm{s}_U)_{a,b}=1$ for $a<b$ and $(\bm{s}_U)_{a,b}=-1$ for $a>b$,
independently of the actual sites contained in $U$.
Then, denoting this $2p\times2p$ matrix $\bm{s}_U$ by $\bm{s}_{2p}$
and expanding its Pfaffian along the first row gives
$
 \Pf \bm{s}_{2p}
 =
 \sum_{j=2}^{2p} (-1)^j \Pf \bm{s}_{2p-2}
 =
 (1-1+\cdots+1-1+1) \Pf \bm{s}_{2p-2}
 =
 \Pf \bm{s}_{2p-2}
$.
Together with $\Pf \bm{s}_2=1$ and the convention $\Pf \bm{s}_0=1$,
this equation proves $\Pf \bm{s}_U=1$ for every up-spin configuration complementary to $D$.
Therefore,
\begin{align}
 \label{eq:identity}
 \Phi(x)
 =\Pf \bm{s}_U \cdot \Pf \bm{\psi}_D
 =1\cdot\Pf \bm{\psi}_D
 =\Psi_{\mathrm{JW}}(D)
\end{align}
holds for any even-sized down-spin configuration $D$.
Note that, for an odd number of down spins, both amplitudes vanish because both states contain only fermion pairs.
Since the proof uses only the Pfaffian form in Eq.~(\ref{eq:jwpf}),
the construction applies to any pure even-parity spinless fermionic Gaussian state on an even number of sites
that has a nonzero overlap with the particle vacuum
and admits the paired exponential representation of Eq.~(\ref{eq:jwbcs}) in the chosen particle basis.

The pairing matrices $F_{i,j}^{\uparrow,\uparrow}$ and $F_{i,j}^{\downarrow,\downarrow}$
in Eq.~(\ref{eq:choice}) are long-ranged in real space,
but both can be obtained from local parent Bogoliubov--de Gennes Hamiltonians on a finite-size antiperiodic momentum grid.
The physical block can be obtained by a nearest-neighbor spinless $p$-wave model
with the hopping amplitude $t_{\alpha}$, the chemical potential $\mu_{\alpha}$,
and the pairing amplitude $\Delta_{\alpha}$, specified by
\begin{align}
 \xi_{k}^{\alpha} = -2t_{\alpha} \cos k - \mu_{\alpha}, \quad
 \Delta_{k}^{\alpha} = 2\ii \Delta_{\alpha} \sin k,
\end{align}
through $F_{k}^{\alpha} = \Delta_{k}^{\alpha}/[E_{k}^{\alpha}+\xi_{k}^{\alpha}]$
with $E_{k}^{\alpha} = \sqrt{(\xi_{k}^{\alpha})^2+|\Delta_{k}^{\alpha}|^2}$ ($\alpha=\uparrow,\downarrow$).
For the spin-down component, we can choose $t_{\downarrow}=\Delta_{\downarrow}=J/2$ and $\mu_{\downarrow}=-Jg$,
which gives the dispersion $E_{k}^{\downarrow}=J\epsilon_k$ of the 1D TFI model.
For the spin-up component, we can choose $t_{\uparrow}=-\Delta_{\uparrow}=J/2$ and $\mu_{\uparrow}=0$,
which results in a flat-band nearest-neighbor Kitaev chain with $E_{k}^{\uparrow}=J$ and $F_{k}^{\uparrow} = s_k = -\ii \cot(k/2)$.
Since $e^{-\ii kL}=-1$ for the antiperiodic momenta and hence $\sum_{r=1}^{L-1}e^{-\ii kr}=-\ii\cot(k/2)$,
the inverse discrete Fourier transform gives $F^{\uparrow,\uparrow}_{i,j} = s_{i,j} = -\operatorname{sgn}(i-j)$
with the antiperiodic continuation $s_{i,j+L}=-s_{i,j}$.

The present construction also applies to even-length spin-$1/2$ chains with JW ground states of the paired Gaussian form in Eq.~(\ref{eq:jwbcs}),
beyond the TFI model.
Examples include
appropriate even-parity ground-state sectors of
the anisotropic XY chain~\cite{lieb1961,barouch1971} and cluster-Ising chains~\cite{smacchia2011,son2012}.
For the latter class, the construction embeds the ground states described by Eq.~(\ref{eq:jwbcs}),
including representatives of the Ising-ordered and cluster-symmetry-protected-topological (SPT) regimes~\cite{chen2011,pollmann2012}.
Therefore, it places the corresponding ordered, critical, and SPT states within a common projected BCS manifold.

\begin{figure}[!t]
\centering
\includegraphics[width=\columnwidth]{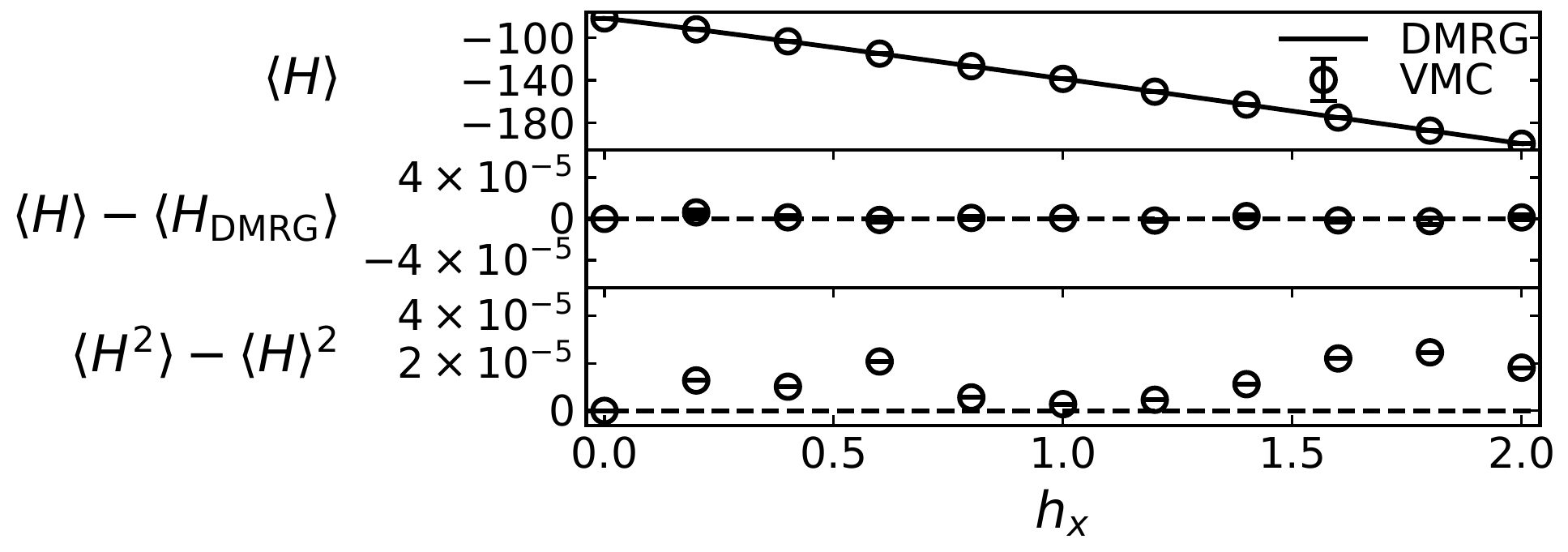}
\caption{Longitudinal-field $h_x$ dependence of physical quantities in the mixed-field Ising model under periodic boundary conditions.
For $L=64$ and $J=h=1$,
the solid line shows nearly exact density matrix renormalization group (DMRG)~\cite{white1992,schollwock2011} results
obtained with the ITensor library~\cite{hauschild2018},
whereas the circles show optimized VMC results obtained with the mVMC library~\cite{tahara2008,misawa2019}.
We show the total energy $\langle H\rangle$ (top),
the difference in total energy $\langle H\rangle - \langle H_{\mathrm{DMRG}}\rangle$ (middle),
and the total energy variance $\langle H^2\rangle - \langle H\rangle^2$ of the optimized VMC state (bottom).
The VMC parameters are optimized using a stochastic reconfiguration scheme with $128000$ samples,
and the statistical errors are estimated by binning.
The DMRG calculations are performed with a maximum bond dimension of $1200$ and a truncation cutoff of $10^{-12}$.
We confirm that the total energy variance of the DMRG ground state is below $10^{-8}$,
at least three orders of magnitude smaller than the VMC value; it is therefore negligible.
The VMC state is exact at $h_x=0$ and the optimized VMC state is very accurate for $h_x>0$.
}
\label{fig:vmc_ene}
\end{figure}

The exact state can also serve as a controlled variational initial state away from integrability.
For example, adding a longitudinal field yields $H=H_{\mathrm{TFI}}-h_x\sum_j\sigma_j^x$,
whose JW representation is no longer quadratic when $h_x\ne0$.
On the other hand,
for $h_x \gg J,h$, the ground state approaches $\prod_j(c_{j,\uparrow}^{\dagger}+c_{j,\downarrow}^{\dagger})\ketvac$ up to normalization.
In general, a product state $\prod_j(u c_{j,\uparrow}^{\dagger}+v c_{j,\downarrow}^{\dagger})\ketvac$ is obtained
by choosing $F_{i,j}^{\alpha,\beta}=s_{i,j}w_{\alpha}w_{\beta}$, where $w_{\uparrow}=u$ and $w_{\downarrow}=v$.
Therefore, we naturally expect the projected BCS state to provide a good variational ansatz for finite $h_x$.
To examine this expectation, we perform variational Monte Carlo (VMC) calculations.
We fix the parameters $J=h=1$ and vary the longitudinal field from $h_x=0$ to $h_x=2$ at $L=64$.
The pairing amplitudes $F_{i,j}^{\alpha,\beta}$ are assumed to be translationally invariant up to an overall sign
and are optimized starting from the $h_x=0$ state specified by $F_{i,j}^{\alpha,\alpha}$,
together with small initial random values of $F_{i,j}^{\uparrow,\downarrow}=-F_{j,i}^{\downarrow,\uparrow}$.
As shown in Fig.~\ref{fig:vmc_ene}, the optimized VMC energy agrees with the nearly exact DMRG energy within the statistical errors.
The total energy variance of the optimized VMC state is less than $3\times 10^{-5}$ for $0\le h_x\le 2$.

In summary, using the exact solution of the 1D TFI model as an example,
we have presented a method for embedding an even-parity spinless Gaussian state
that admits a paired exponential representation on an even lattice
into an Abrikosov-fermion projected BCS state.
In this embedding, the auxiliary up-spin Pfaffian is unity for any choice of up-spin sites,
yielding an exact projected BCS representation of the Gaussian state in the physical spin Hilbert space.
As a complement to matrix-product-state-based tensor-network methods~\cite{jin2022},
our exact projected BCS embedding provides a natural starting point for variational studies of nonintegrable systems,
potentially including frustration-free models with exact Gaussian ground states~\cite{peschel1981,sierant2026}.
Extensions to odd-parity states, multiband Gaussian states, and higher-dimensional fermionization with nontrivial gauge sectors remain open.

\begin{acknowledgments}
This work was financially supported by
MEXT KAKENHI, Grant-in-Aid for Transformative Research Areas
(Grant Nos.\ JP22H05111 and JP22H05114)
and
JSPS KAKENHI
(Grant Nos.\ JP25K07157 and JP26K06959).
The numerical computations were performed on computers at
the Supercomputer Center, Institute for Solid State Physics, University of Tokyo.
\end{acknowledgments}

\balance
\bibliographystyle{jpsjnum}
\bibliography{references}

\end{document}